\documentclass[review]{elsarticle}
\usepackage{graphicx}
\usepackage{enumerate}
\usepackage{longtable}
\usepackage{float}
\journal{Fusion Engineering and Design}
\begin{document}

\begin{frontmatter}

\title{RMATE: A device to test radiation-induced effects under controlled magnetic field and temperature}

\author[CIEMAT]{I.~Garc\'{i}a-Cort\'{e}s}
\author[CIEMAT]{S.~Cabrera}
\author[CIEMAT]{M.~Medrano}
\author[CIEMAT]{A.~Moro\~{n}o}
\author[CIEMAT]{P.~Mu\~{n}oz}
\author[CIEMAT]{A.~Soleto}
\author[SEGAINVEX]{I.~Bugallo}
\author[IMA]{A.~Nieto}
\author[IMA]{R.~Altimira}
\author[IMDEA]{A.~Bollero}
\author[UAM,IMDEA]{J.~Camarero}
\author[IMDEA]{J.~L.~F.~Cu\~{n}ado}
\ead{joseluis.fcunnado@imdea.es}
\address[CIEMAT]{Laboratorio Nacional de Fusi\'{o}n, CIEMAT. Avda. Complutense 40, 28040 Madrid, Spain}
\address[SEGAINVEX]{Servicios de Apoyo a la Investigaci\'{o}n, SEGAINVEX-UAM,Universidad Aut\'{o}noma de Madrid, 28049 Madrid, Spain}
\address[IMA]{Ingenier\'{i}a Magn\'{e}tica Aplicada, Avda. Catalunya, 5, 08291 Ripollet – Barcelona, Spain }
\address[IMDEA]{Instituto Madrile\~{n}o de Estudios Avanzados en Nanociencia IMDEA-Nanociencia, Campus Universidad Aut\'{o}noma de Madrid, 28049 Madrid, Spain}
\address[UAM]{Departamento de F\'{i}sica de la Materia Condensada and Instituto "Nicol\'as Cabrera", Universidad Aut\'{o}noma de Madrid, 28049 Madrid, Spain}

\begin{abstract}
This study shows the development and performance assessment of a novel set-up that enables the research of structural materials for fusion reactors, by making possible simultaneous application of temperature (up to 450$^{\circ}$C) and magnetic field (close to 0.6~T) during irradiation experiments. These aspects become critical as structural materials in fusion reactors are exposed to intense radiation levels under the presence of strong magnetic fields. Moreover, material micro-structural could be modified by radiation-induce propagating defects, which are thought to be sensitive to magnetic fields. The device has three main components: magnetic closure, sample holder with integrated heater, and radiation shield. It is provided with a thermal shield to prevent other elements of the device to heat up and fail. A mapping of the magnetic flux in the region where sample and heater are located has been modeled by finite elements simulation software and correlated with magnetic measurements.
\end{abstract}


\end{frontmatter}

\section{Introduction}
Fusion reactors based on magnetic confinement need special materials for structural and vacuum vessel, capable of resisting high level of radiation in presence of the strong magnetic fields required for plasma confinement~\cite{2015__NF_____MOTOJIMA,2014__JNM____STORK_AGOSTINI}. In consequence, the selection and design of the materials that can face the extreme conditions of fusion power plant have been described as one of the great material science challenges~\cite{2017__NF_____STORK_ZINKLE, 2005__PP_____ZINKLE}. High-chromium ferritic/martensitic steels are good candidates as structural materials for fusion reactors because of their high resistance to irradiation~\cite{1979__JNM____LITTLE_STOW, 2000__JNM____GARNER_TOLOCZKO}. However, it is well-known that micro-structural and mechanical properties of these materials are modified by radiation-induce propagating defects~\cite{2016__NME____SANCHEZ_GARCIA-CORTES,2014__PRB____GOMEZ-FERRER_GARCIA-CORTES}. In addition, since they are based on ~{Fe-Cr} alloys, which are ferromagnetic materials, the presence of high magnetic fields could also play a role in the performance of these steels. Concerning this subject, several theoretical studies point to magnetism as being a critical factor in radiation induced damage in structural materials~\cite{2005__PRL____SELETSKAIA_OSETSKY, 2008__JNM____MALERBA_CARO}. These predictions suggest that magnetism can be a non-negligible factor in defining the defect properties induced by ion irradiation or in determining the atomic distribution. However, a few experimental and theoretical efforts have been done in order to elucidate the role played by magnetism in damage. It has been found that Fe$^{+}$ irradiation on ferromagnetic films induced changes in the magnetic state for various irradiation doses~\cite{2016__NME____PAPAMIHAIL_MERGIA_OTT, 2016__PRB____PAPAMIHAIL_MERGIA_OTT}. 

In this context, recent irradiation experiments in ~{Fe-Cr} alloys  were carried out to study the influence of external magnetic fields in damage under field strengths of around 0.4~T at room temperature  ~\cite{2016__NME____SANCHEZ_GARCIA-CORTES, 2019__JNM____GARCIA-CORTES_LEGUEY}. The results of these works point to the external magnetic field as a non-negligible parameter in vacancy profile and chromium mobility due to the defect creation. However permanent magnets used in these works have magnetic field values well below of those expected in future reactors (several Tesla). On another hand, the approach used thus far to set the external field on the sample was to fit it in contact with the permanent magnet, where the magnetic field is maximum, but this means a drawback in order to heat the sample, as this would alter the characteristics, or even damage the permanent magnet by going above its maximum operation temperature. Accordingly the experimental set up of these works does not allow the realization of experiments at high temperature (above 200$^{\circ}$C) with application of a maximum magnetic field. A direct consequence of this limitation is the scarcity of results on the behavior of structural materials under relevant conditions for fusion reactors applications, i.e., high temperature and high magnetic fields.

In order to overcome these limitations, a new device, named RMATE (for \emph{Radiation-induced effects under controlled MAgnetic field and TEmperature}), has been designed and construct from scratch. It consists of a closed chamber in whose interior the sample can be located, exposed to the ion beam through an axial entrance, at controlled conditions of temperature and magnetic field, solving the aforementioned problems. The chamber itself constitutes a magnetic closure that drives the magnetic flux and concentrates it at the sample position. Flux lines are kept parallel to the ion beam, so that deflection due to Lorentz force is avoided. RMATE has been designed to be mounted inside one of the vacuum chambers of the Danfysik implanter line, at the research center CIEMAT (Madrid, Spain)~\cite{2017__JNM____MUNOZ_GARCIA-CORTES,2018__JNM____MUNOZ_GARCIA-CORTES}. For this reason, its shape has constrains imposed by the implanter chamber in which is located, and under this restrictions, the shape has been optimized to get the maximum field values at the sample position. This manuscript is organized as follows: a description of the device is introduced first, followed by simulation of  the magnetic flux in the magnetic closure based on optimized shape; each element of the device, as finally built in our workshops, is described accompanied by magnetic characterization. The study concludes with a discussion of the results as well as discussion of future improvement perspectives.

\section{Design of the irradiation device prototype}

\subsection{General description} 
\begin{figure}[h]
\begin{center}
\includegraphics*[width=75mm]{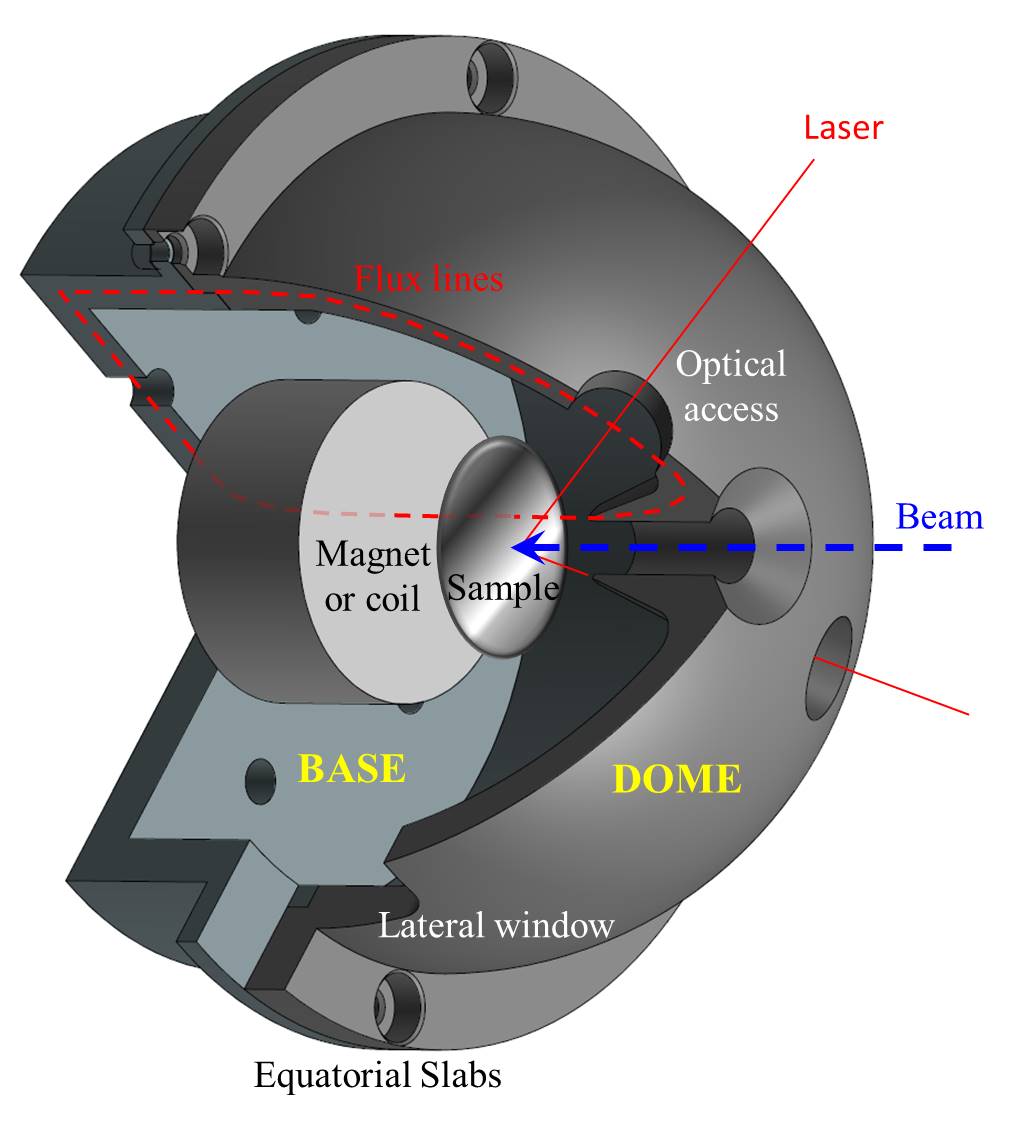}
\caption [Figure1]{\label{fig:Figure1}
3D view, with quarter cutoff, of the magnetic closure. The two parts of the revolution-shaped chamber can be identified (indicated as BASE and DOME). The dome is provided with the beam access (dashed blue arrow). Magnetic flux lines follow a closed circuit, as depicted by the dashed red line (visual guide), running through the axis and the perimeter. The magnet can be seen here sitting on the base, and the sample is represented as a disk in between the magnet and the entrance cone. Two oblique optical accesses have been set to permit in-situ vectorial magneto-optical Kerr effect magnetometry (v-MOKE), using laser light, represented here by a red line reflection on the sample surface.
}
\end{center}
\end{figure}

The RMATE device is designed to keep the sample at temperature of up to 450$^{\circ}$C under magnetic fields of 0.3-0.6~T (with a single permanent magnet and depending of the ferromagnetic nature of the sample under study), allowing at the same time to expose the sample to ion beam irradiation. The magnetic field can be created either by a permanent magnet or a coil inside the magnetic closure (in this paper we focus in the simplest case of a permanent magnet), while the temperature is controlled by using an oven or a filament heater. Three main components compose the device: the magnetic closure, the sample holder with an integrated heater and a tandem radiation shield. The sample is set inside the magnetic closure, at its axis, facing an entrance through which it is exposed to the ion beam. The magnetic closure has been conceived as a closed cavity generated by a revolution profile, as show in the 3D scheme of Figure~\ref{fig:Figure1}, where most of the inner parts have been removed for clarity. The closure splits into two parts, (which are joint with screws at their equator's slabs): a \emph{base} part ending in a flat face (left part of Figure~\ref{fig:Figure1}), and the so-called \emph{dome}, due to its particular shape, (right part of Figure~\ref{fig:Figure1}). The beam entrance is an inward perforated cone located in the dome, at its axis, as indicated in the figure. A NdFeB permanent magnet (N35 type, 30~mm diameter and 10~mm thick) is axially located in the base, sitting on its inner side. The particular closed \emph{`pot-like'} shape of the magnetic closure, made of ferromagnetic material, forces magnetic lines to concentrate in the axis of the chamber, closing magnetic flux through its perimeter, as indicated by the dashed red closed line depicted in the upper cut (visual schematic illustration). The final profile of the revolution figure is the result of an optimization process carried out by simulation, which will be detailed in the following sections.

\subsection{Magnetic flux simulation} 
\begin{figure}[h]
\begin{center}
\includegraphics*[width=85mm]{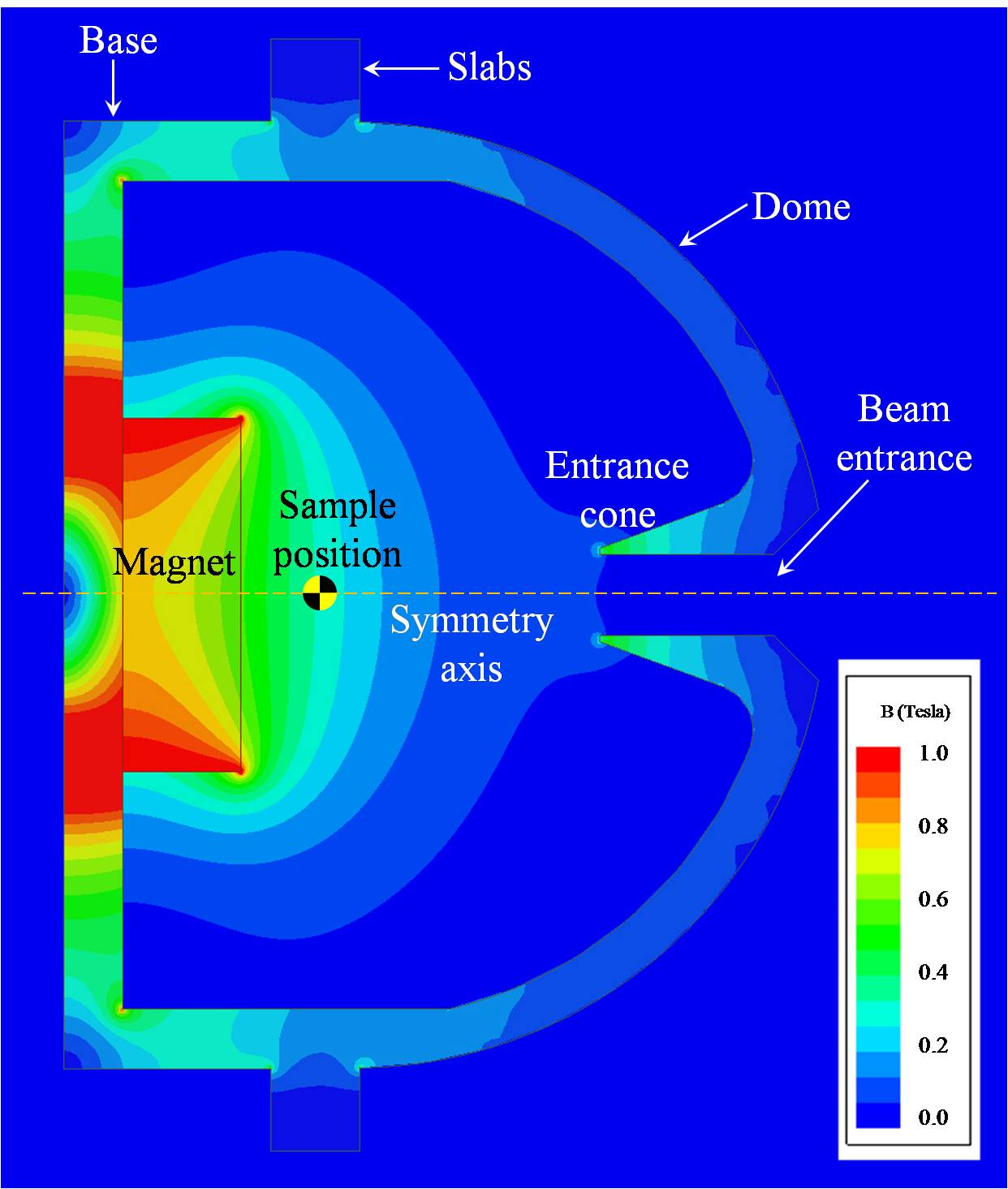}
\caption [Figure2]{\label{fig:Figure2}
Simulation of the prototype magnetic closure, using a single NdFeB magnet. Color pattern is used to identify the magnetic field intensity (in Tesla), from highest (red) to lowest intensities (dark blue). These simulations have been done on the basis of AISI 420 steel. The sample is located at the position indicated by the yellow and black bladed disk. Magnetic flux at such position is about to 0.4~T. 
}
\end{center}
\end{figure}

\begin{figure}[h]
\begin{center}
\includegraphics*[width=85mm]{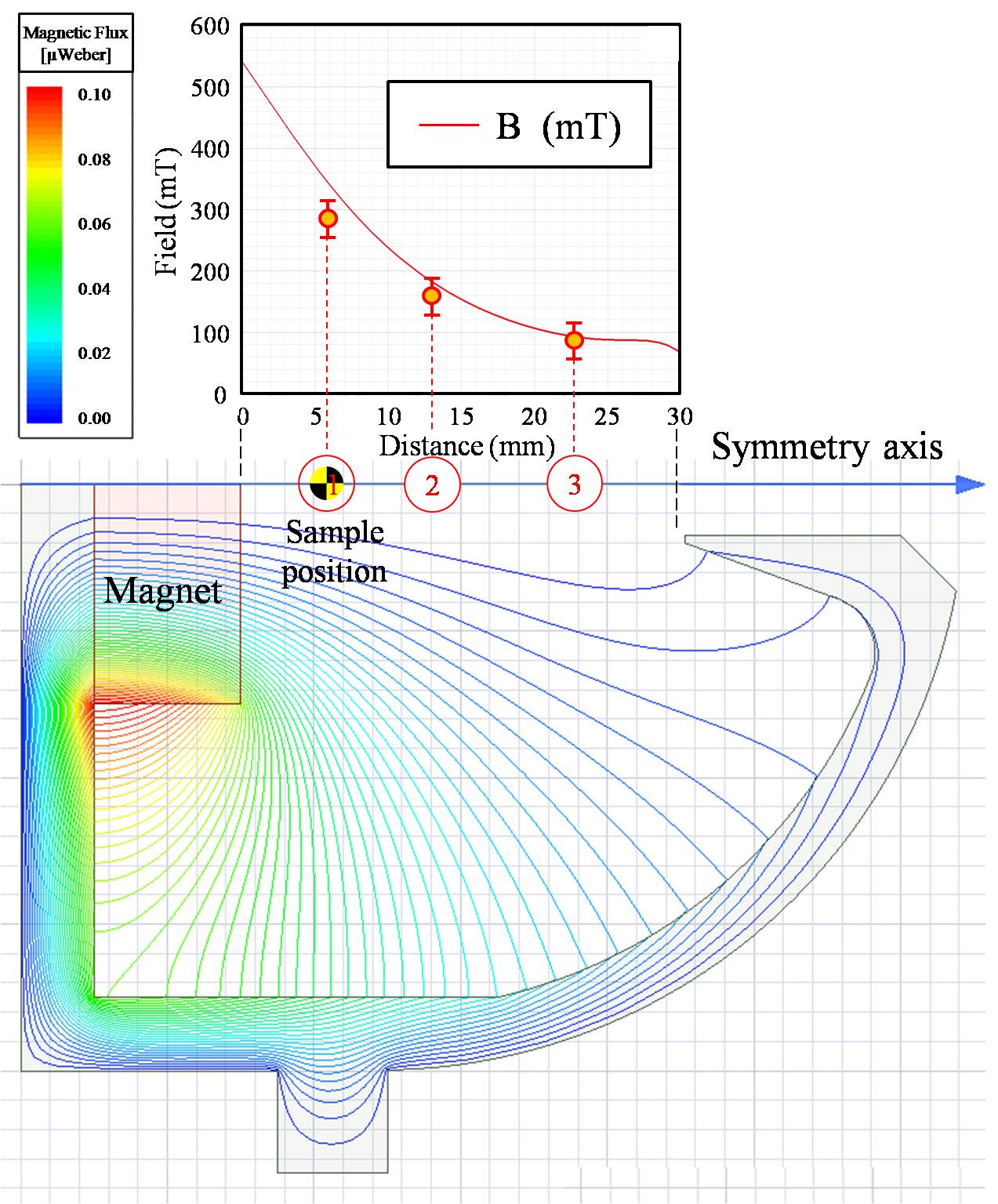}
\caption [Figure3]{\label{fig:Figure3}
Magnetic flux simulation for the prototype can be seen in color line scheme. A magnetic field profile taken at the axis is displayed on top. This profile corresponds to the region between the magnet surface (set as 0~mm in the horizontal axis) and the entrance cone (30~mm away). Three points where measured experimentally, at positions marked in red circles (see the text and Figure~\ref{fig:Figure6}). The corresponding experimental data can be seen superimposed on the field profile, with red symbols (yellow filled). Reference dashed red lines have been depicted for clarity.
}
\end{center}
\end{figure}

In order to create the profile shape of the magnetic closure, we first started with an approximate initial shape and carried out sequential simulations until we got to the final (optimized) shape. The initial shape  consisted of a rough revolution figure with the magnet in the inner side of its base, and the beam entrance at its opposite side (early-stage pot-shaped closure). This initial shape was designed to fit inside the Danfysik ion implanter chamber. Finite elements simulation software FEMM~\cite{Ref_FEMM} and MAXWELL~\cite{Ref_ANSYS_MAXWELL} where used to create the final optimized magnetic closure profile, as can be seen in Figure~\ref{fig:Figure2}. The optimization pursues to get the maximum magnetic field in the sample position, located at the axis inside of the magnetic closure, while keeping flux lines parallel to it, so that the ion beam is not deflected by Lorentz force. This ensures the beam reaching the sample without any loss of intensity and/or energy. Figure~\ref{fig:Figure2} shows the magnetic field intensity simulation as a color pattern superimposed to the profile of the closure cross section. The material used in the simulation is AISI 420 steel. The magnetic field intensity color pattern goes from red for the highest field values to blue for the lowest (dark blue being zero, according to color code displayed in the legend of the figure). The magnet can be seen sitting on the inner base side, on the left, while the perforated entrance cone can be seen in the dome at the right. Different tests where made at Ingenier\'{i}a Magn\'{e}tica Aplicada (IMA)~\cite{Ref_IMA} for several shapes, in order to reach an optimized one. The sample position is schematically indicated in Figure~\ref{fig:Figure2} as a small disk with yellow and black blades, located between the magnet and the beam entrance. In this point, simulations show a field value around 0.4 T. In Figure~\ref{fig:Figure3} the magnetic flux lines can be seen, together with a plot of the magnetic field as a function of the distance from the magnet, in direction of the entrance cone. 

\section{Description of the device elements}

\subsection{Magnetic closure}
\begin{figure}[h]
\begin{center}
\includegraphics*[width=85mm]{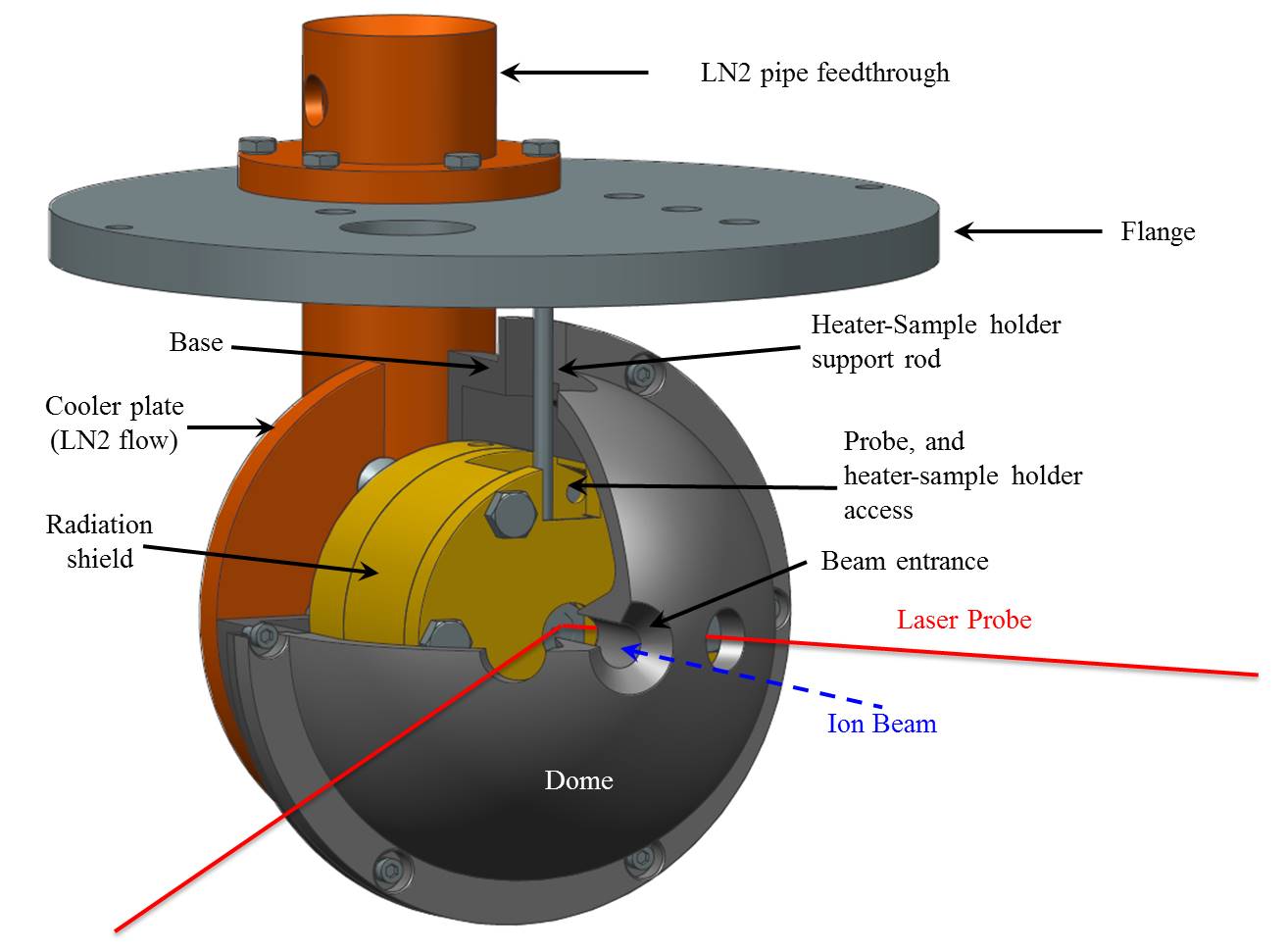}
\caption [Figure4]{\label{fig:Figure4}
3D scheme of the magnetic closure as mounted on the flange for the Danfysik irradiation chamber. The closure appears with a quarter cutoff that allows to see the tandem radiation shield. The cooler plate can be seen behind the closure in orange color, attached to the cold finger, which is the vertical orange cylinder. The cold finger acts both as supporting structure and feedthrough. The aperture in the radiation shield can be seen, as well as the laser, used for in-situ v-MOKE magnetometry (indicated by the red line) and the lateral optical accesses. A rod emerges from inside the radiation shield through a window, traversing also the magnetic closure through another window, ending at the flange, where is fixed. This rod supports the heater-sample holder (not seen here, refer to Figure~\ref{fig:Figure5}) inside the radiation shield, without touching any part of it.
}
\end{center}
\end{figure}

For this first prototype of the magnetic closure we used AISI 420 RE steel, easy to acquire and mechanize. It consists in an axis-symmetric hollow pot divided at its equator into the already described base and dome parts, fixed together by screws located in slabs at the joint. 3D schematics of the design can be seen in Figure~\ref{fig:Figure1}, where the revolution shape has been cutoff in a quarter view. The two parts are clearly identified: 
\begin{enumerate}
\item Base: this is the left part of the closure shown in Figure~\ref{fig:Figure1}, where the whole device is supported. The axial magnet is located in the inner side of the base, represented  with a cylinder in the figure. Since the base is in contact with the cooler plate, it acts as thermal sink for the rest of the closure. More details are explained later with the help of Figures~\ref{fig:Figure4} and~\ref{fig:Figure5}.

\item Dome: this is the part on the right side of ~\ref{fig:Figure1}. It has the ion beam entrance, that can be seen as an inward perforated cone. Two additional oblique entrances are located symmetrically besides the beam entrance, at a certain angle. These are conceived for future optical access suitable for in-situ vectorial resolved magneto-optical Kerr effect magnetometry (v-MOKE)~\cite{2015__RSI_____Cunado, 2014__RSI_____Cunado}, so that a laser beam can go through one of these oblique accesses to reach easily the sample, and go out after reflecting on it through the other (symmetrically located) oblique access. These oblique accesses can be also seen in Figure~\ref{fig:Figure4}. Between the magnet and the entrance cone, there is room enough to allocate the radiation shield and the sample holder with the heater, in such a way that the sample is decoupled from the magnet and the radiation shield inner walls. This decoupling of the sample holder and heater is carried out by using a support rod directly fixed to the support flange (explained later, see also Figure~\ref{fig:Figure5}).
\end{enumerate}

Schematic illustration in Figure~\ref{fig:Figure1} has been simplified to allow for a better view of the magnetic closure profile, by showing only the magnet inside and the sample disk (a complete cross section can be seen in Figure~\ref{fig:Figure6}). The magnet, radiation shield and heater-sample holder set-up are collinear with the closure axis. A lateral window (cut by the sectioning line) can be seen at the lower part of the figure. This window is used to allow the sample support rod going out to the flange, where it is fixed, and also allows the wires and sensors to get in. Figures~\ref{fig:Figure4} and ~\ref{fig:Figure5} provide more details.  
 
\subsection{Sample holder and heater}
Since the research of materials for fusion reactor applications requires setting the samples at controlled high temperatures, a heater is necessary. RMATE is provided with a sample holder that has an integrated heater on it. In Figure~\ref{fig:Figure5} the sample holder can be observed hanging from the flange by a rod fixed to it with a non-conducting ceramic waffle, not seen in this figure (refer to Figure~\ref{fig:Figure7}a). In this way, we prevent any contact with the rest of the device elements, thus avoiding heat conduction through the structure. The sample is mounted in the inner side of the sample holder, so that it is as close as possible to the magnet. The heat generated in the sample holder is dissipated by the thermal shield. An additional braid (not shown) connects the heater-sample holder rod to the cooler plate. This is important because the rod is fixed to the flange through a ceramic piece and hence specific cooling must be set on it.

\subsection{Radiation shield}
\begin{figure}[h]
\begin{center}
\includegraphics*[width=85mm]{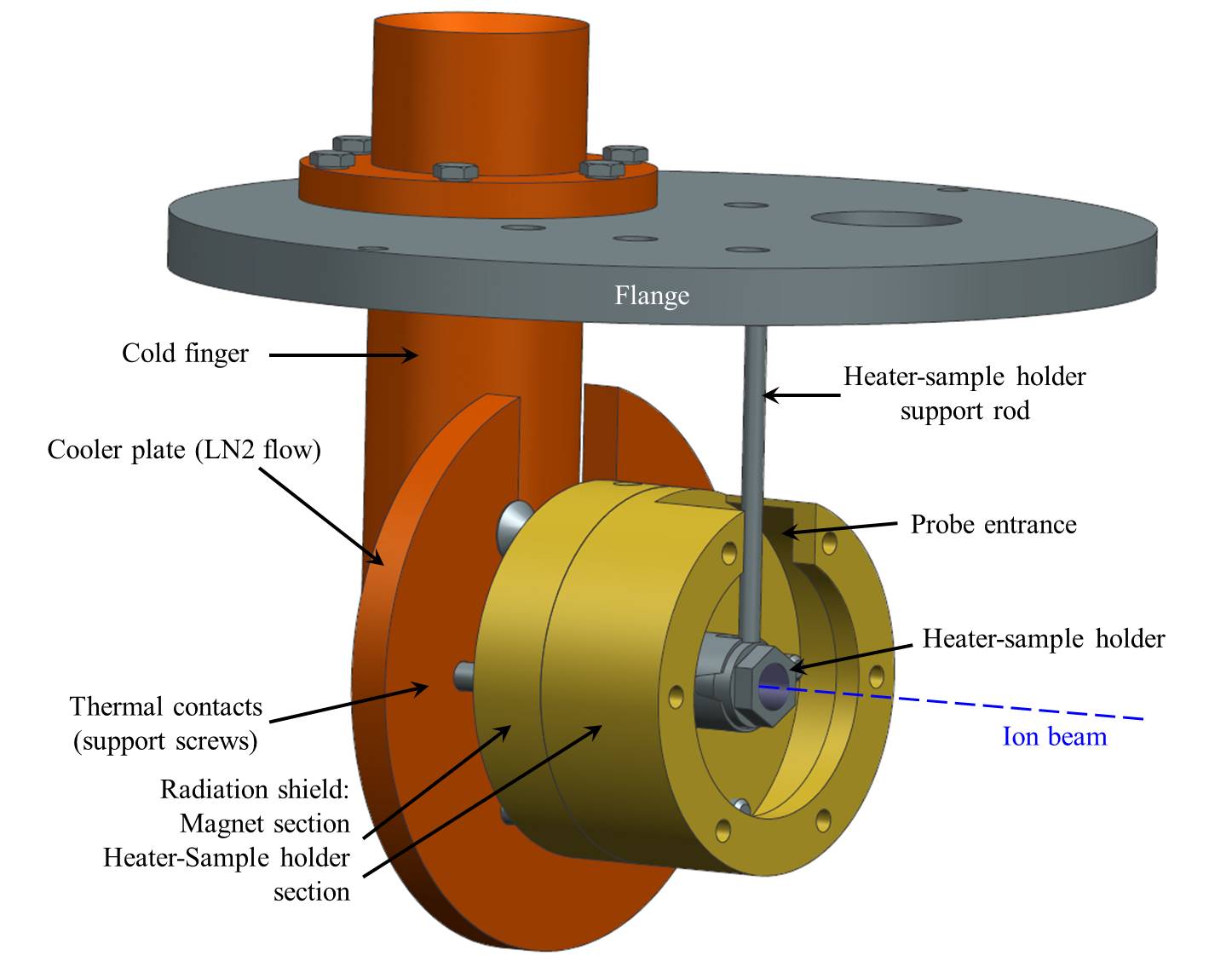}
\caption [Figure5]{\label{fig:Figure5}
Details of the heater-sample holder, its supporting rod, radiation shield, cooler plate and cold finger. The magnetic closure and the radiation shield cap have been remove in this 3D scheme, so that the sample holder and heater can be seen. Thermal contact with the cooler plate can be seen in gray color (although they are copper screws). These have double functionality: they keep magnetic closure and radiation shield fixed to the cooler plate, and allow heat sink from radiation shield and closure into the cooler plate and cold finger. 
}
\end{center}
\end{figure}

Radiation shield becomes necessary in order to prevent overheating of the magnet (NdFeB-based magnets have a maximum working temperature of around 200$^{\circ}$C) or magnetic closure. A tandem radiation shield has been designed so that first, the magnet is thermally separated from the heater-sample holder, and second, the heater does not affect the closure, preventing the magnet from being indirectly heated up through it. The tandem radiation shield can be identified in yellow color in both Figures~\ref{fig:Figure4} and~\ref{fig:Figure5}. The left part of the tandem corresponds to the radiation shield of the magnet itself, indicated as \emph{Magnet section} in Figure~\ref{fig:Figure5},  and the right part corresponds to the heater-sample holder, indicated as \emph{heater-sample holder section} in the same figure. 

\begin{figure}[h]
\begin{center}
\includegraphics*[width=88mm]{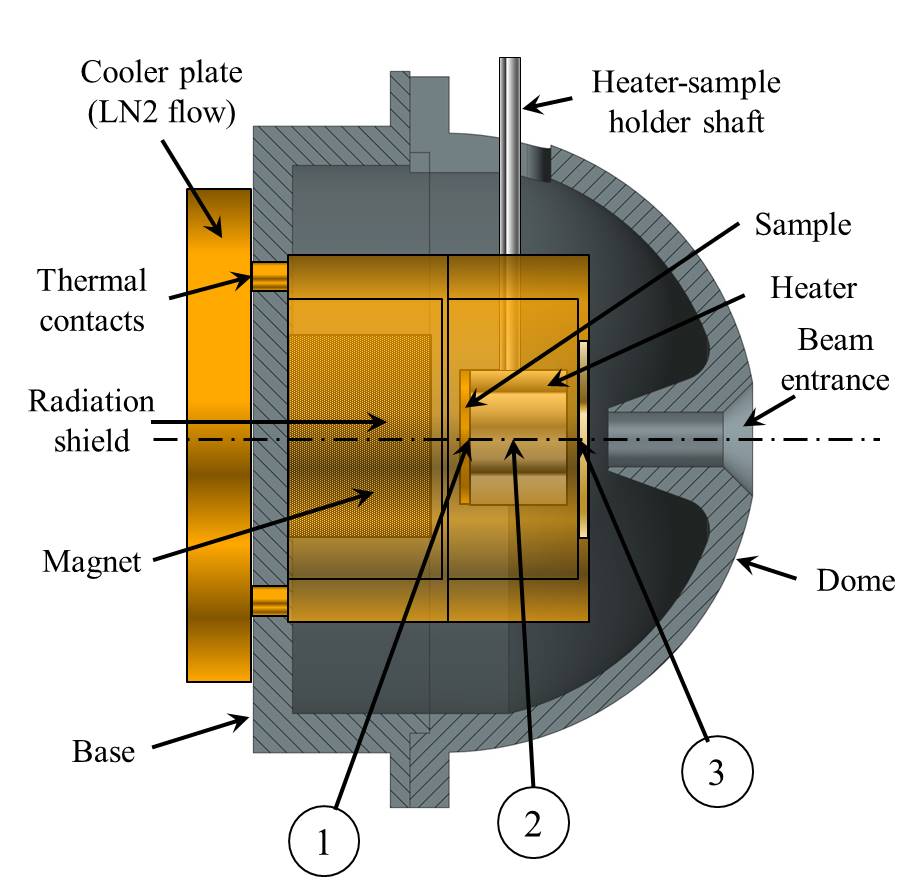}
\caption [Figure6]{\label{fig:Figure6}
Schematic diagram of the cross-section of the magnetic closure with all the inner elements on its actual position. Notice that the sample position is as close as possible to the point where the maximum of magnetic field is obtained. Points 1,2 and 3 indicate the positions where the magnetic field has been experimentally measured in the prototype.
}
\end{center}
\end{figure}

In order to keep the radiation shield at low temperatures, six copper screws (M6) connects it thermally to a cooler plate, traversing the magnetic closure base. This design allows that both the radiation shield and the base are in thermal contact with the cooler plate. In Figure~\ref{fig:Figure4} the magnetic closure appears with a 1/4 cutoff, allowing to see the cooler plate and the radiation shield. In Figure~\ref{fig:Figure5} the magnetic closure has been removed so that the connection between the tandem radiation shield and the cooler plate can be observed (screws appear almost hidden). In Figure~\ref{fig:Figure6}, the whole device cross section is depicted, showing how the cooler plate, the base and the radiation shield are fixed by the six aforementioned screws, ensuring a total thermal contact between them, while the heater-sample holder remains isolated.

The cold finger supports the cooler plate, and it also acts as structural element to support the whole device to the Danfysik chamber flange (Figure~\ref{fig:Figure7}). 

\section{Assembly and characterization of the prototype}

The final prototype of the magnetic closure device was built at UAM-SEGAINVEX and CIEMAT workshops. The magnetic closure was made of AISI 420 RE steel, with a base diameter of 80~mm and a total length of 62~mm. The cooler plate has the same diameter as the base, and it is made of 7~mm thick copper. It is provided with an additional copper thin foil in between, in order to ensure an optimal thermal contact. The cooler plate is fixed to the cold finger, made of copper and also in good thermal contact. 

\begin{figure}[h]
\begin{center}
\includegraphics*[width=75mm]{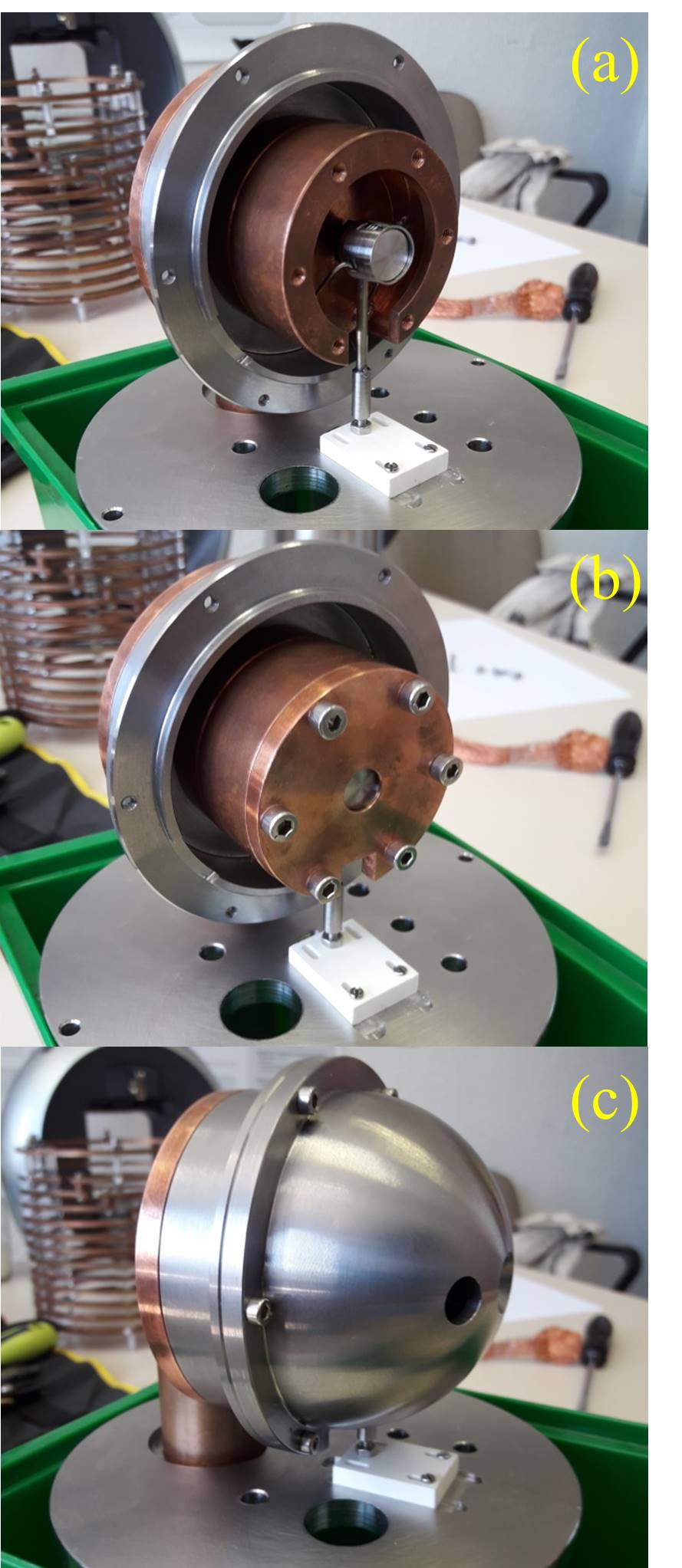}
\caption [Figure7]{\label{fig:Figure7}
Pictures of magnetic closure (explanations in the text).
}
\end{center}
\end{figure}

In Figure~\ref{fig:Figure7} three pictures of the set-up are shown. The set-up is upside down, with the flange here acting as a support platform. The cold finger and cold plate can be seen supporting the device from its base through the cooler plate (better seen in Figure~\ref{fig:Figure7}(b)). In Figure~\ref{fig:Figure7}(a), the dome and the cap of the radiation shield have been removed, allowing  to see the heater-sample holder set-up, and the rod that supports it to the flange. This rod ends in a mechanizable Makor ceramics to isolate electric contacts. Notice the lateral window of the radiation shield through which the heater-sample holder rod and any required electrical contacts can traverse the closure reaching the ceramic support. In Figure~\ref{fig:Figure7}(b), the cap of the radiation shield has already been put on (the screws that fix the cup are independent of the copper screws that fix the radiation shield, the base and the cooler plate). The beam entrance through the cap of the radiation shield is visible here. Notice that the heater-sample holder is not in contact with any part of the radiation shield. The cap of the radiation shield has a hole to allow the ion beam to reach the sample. In Figure~\ref{fig:Figure7}(c), the dome of the magnetic closure has been set back, allowing to see the set-up as it is mounted inside the final vacuum chamber of the CIEMAT Danfysik ion implanter.\\

Magnetic field characterization of the RMATE device was carried by testing field values at three points inside the device, always in the axis, indicated with numbers 1 through 3 in Figures~\ref{fig:Figure3} and~\ref{fig:Figure6}. At position 1, where the sample is located, a value of 298~mT was measured. At position 2, the field drops to 162~mT, and at point 3 value measured is 93~mT. As can be seen in Figure~\ref{fig:Figure3}, top graph, the values are in agreement with those expected by the simulations. Notice that because of size and shape restrictions, the shape of the dome makes flux lines to scape towards the dome walls, rather than towards the entrance cone. This is due to proximity and reduces concentration in the axis. 

\begin{figure}[h]
\begin{center}
\includegraphics*[width=85mm]{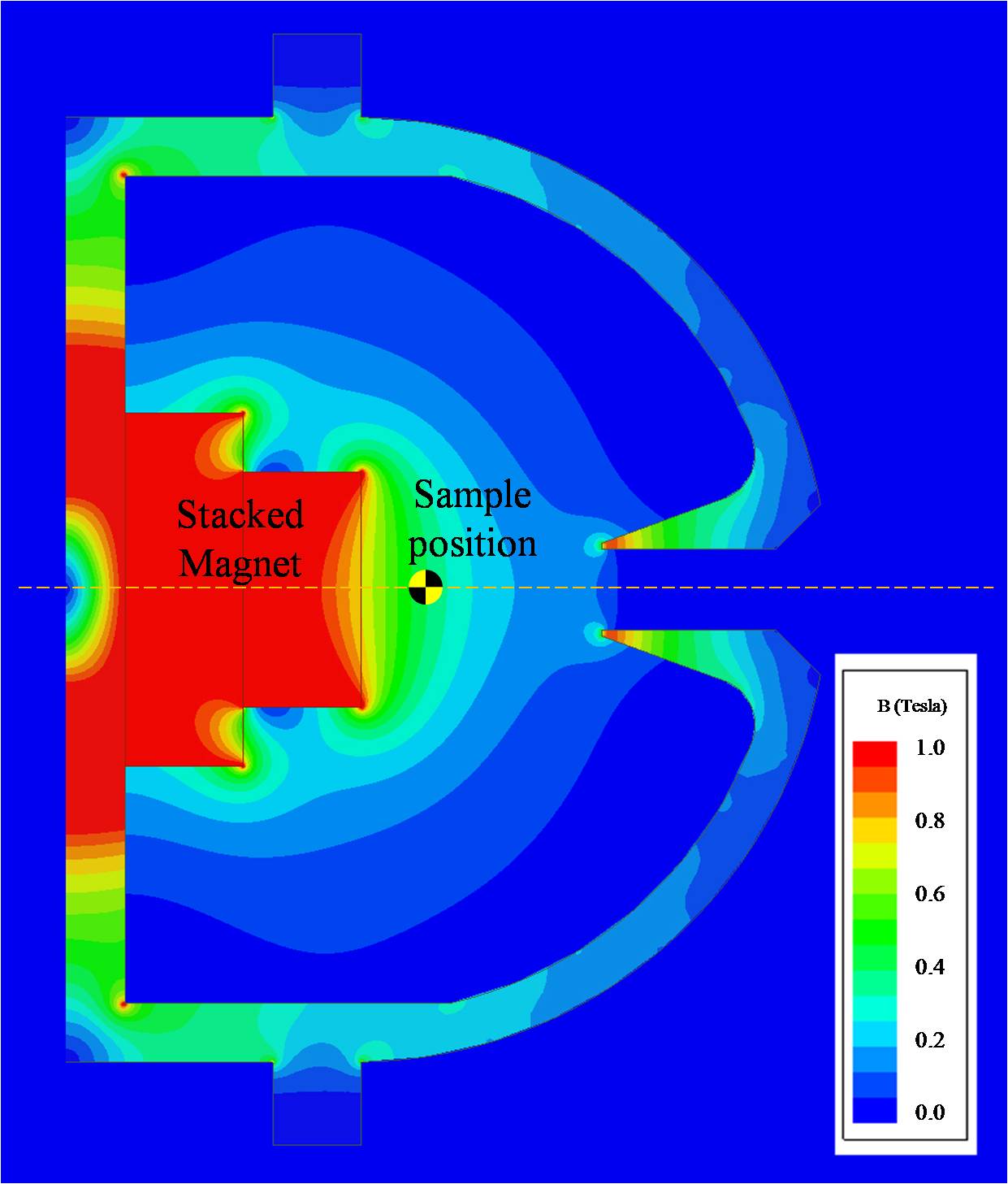}
\caption [Figure8]{\label{fig:Figure8}
Simulation of the magnetic closure with stacked magnet can be seen here. At the indicated sample position, fields of about 0.6~T can be reached. When a ferromagnetic sample is located in that position, re-conducted magnetic flux lines allows reaching higher fields, close to 1~T.
}
\end{center}
\end{figure}

In order to overcome this issue and reach higher field values at sample position, there are two possibilities (that could also be combined): on the one hand, a stacked-magnet configuration could be used to concentrate the flux lines closer to the axis; on the other hand, a ferromagnetic sample holder with particular design can be used to attract magnetic flux to the sample position. Stacked-Magnet configuration has been simulated, and can be seen in Figure~\ref{fig:Figure8}. In this configuration, the field value at the expected sample position (red dashed cross  in the figure) will be around 0.55~T. This configuration can be mounted by simply reconfiguring the second section of the radiation shield, provided that the heater is slightly reduced in length. Simulations where carried out also with a sample holder consisting on a ferromagnetic AISI 420 disk (not shown). The field thus obtained could reach values as high as 0.8~to 1~T. Notice that, as the samples under research for fusion applications itself is ferromagnetic, the final field that it sees inside the magnetic closer will be higher than the measured one for our prototype. Also it must be taken into account that the steel used in the prototype is not the ideal one, and further simulations with better steels are now underway, together with more tests to improve the closure profile. We expect to get field values close to 1~T in near future. In any case, the presented prototype already allows for reasonable magnetic field values at sample position, under controlled temperatures and with the possibility of ins-situ v-MOKE magnetometry. 

\section{Conclusions}
This is the first time that on purpose magnetic closure/sample holder device is built for the investigation of temperature and field-dependent irradiation damage of structural materials for fusion reactors. It has been developed to be fitted in the ion implanter (Danfysik) irradiation facility currently in operation at CIEMAT. This compact irradiation device is very easy to adapt to bigger implanter. The magnetic closure profile has been optimized to concentrate magnetic flux in the sample position, reaching values of the order of 0.3~T or higher when ferromagnetic sample is mounted. It can also keep the sample under controlled temperatures of up to 450$^{\circ}$C, and, remarkably, its design permits to carry out in-situ vectorial resolve magneto-optical Kerr effect magnetometry. The magnetic closure can be further optimized by improving constituent materials, or even more, modifying the shape in case space restrictions are not present (as is the case in other ion implanter facilities), and opens new perspectives in the research of materials for fusion reactor applications.  

\section{Acknowledges}
This work was mainly financed by the following projects: Spanish MINECO (Ministerio de Econom\'{i}a y Competitividad) under project ENE2016-76755-R (AEI/FEDER, UE), Comunidad de Madrid through projects NANOMAGCOST-CM, project n.~P2018/NMT4321, and NANOFRONTMAG, project n.~S2013/MIT-2850. P.~M. acknowledges a pre-PhD contract of the Spanish MINECO. IMDEA nanociencia acknowledges support from the Severo Ochoa Program (MINECO, Grant SEV-2016-0686). We also acknowledge support from UAM-SEGAINVEX, and in particular, we acknowledge to J. R. Marijuan and J.M. Cortés.

\end{document}